\documentclass[prd,preprint,tightenlines,floatfix,showpacs,preprintnumbers,nofootinbib,eqsecnum]{revtex4}
 \usepackage[dvips,final]{graphicx}
  \usepackage{amssymb}     \usepackage{amsmath}
    \usepackage{amsfonts}       \usepackage{epsfig}
      \usepackage{bm}

\def\GeV{\text{{\rm GeV}}}  \def\MeV{\text{{\rm MeV}}}

\newcommand{\ac}{\mathcal{A}} 
\newcommand{\beq}{\begin{equation}} \newcommand{\eeq}{\end{equation}}

\def\al{\relax\ifmmode\alpha\else{$\alpha${ }}\fi}
\def\alps{\relax\ifmmode\alpha_s\else{$\alpha_s${ }}\fi}
\def\as{\relax\ifmmode\alpha_s\else{$\alpha_s${ }}\fi}
\def\msbar{\relax\ifmmode\overline{\rm MS}\else{$\overline{\rm MS}${ }}\fi}
\def\albar{\relax\ifmmode{\bar{\alpha}}\else{$\bar{\alpha}${ }}\fi}
\def\albarE{\relax\ifmmode{\bar{\alpha}_E}\else{$\bar{\alpha}_E${ }}\fi}
\def\alphaE{\relax\ifmmode{\alpha_E}\else{$\alpha_E${ }}\fi}
\def\albarEQ{\relax\ifmmode{\albar_E(Q^2)}\else{$\albar_E(Q^2)${ }}\fi}
\def\albarM{\relax\ifmmode{\bar{\alpha}_M}\else{$\bar{\alpha}_M${ }}\fi}
\def\albars{\relax\ifmmode{\bar{\alpha}_s}\else{$\bar{\alpha}_s${ }}\fi}
\def\albarsQ{\relax\ifmmode{\bar{\alpha}_s(Q^2)}\else{$\,\bar{\alpha}_s(Q^2)${}}\fi}
\def\agoth{\relax\ifmmode{\mathfrak A}\else{$\,{\mathfrak A}${ }}\fi}
 \def\agothk{\relax\ifmmode{\mathfrak A}_k\else{${\mathfrak A}_k${ }}\fi}
\def\agothks{\relax\ifmmode{\mathfrak A}_k(s)\else{${\mathfrak A}_k(s)${}}\fi}
\def\acal{\relax\ifmmode{\cal A}\else{${\cal A}${ }}\fi}
  \def\acalk{\relax\ifmmode{\cal A}_k\else{${\cal A}_k${ }}\fi}
 \def\acalkQ{\relax\ifmmode{\cal A}_k(Q^2)\else{${\cal A}_k(Q^2)${ }}\fi}
\def\alphaMs{\relax\ifmmode\alpha_M(s)\else{$\alpha_M(s)${ }}\fi}
\def\alphaEQ{\relax\ifmmode{\alpha_E(Q^2)}\else{$\alpha_E(Q^2)${ }}\fi}
\def\alphaM{\relax\ifmmode{\alpha}_M\else{$\alpha_M${ }}\fi}

\graphicspath{./BAPTfigs/}

\begin{document}
\thispagestyle{empty} \preprint{\hbox{}} \vspace*{-10mm}

\title{Nucleon spin structure and pQCD frontier on the move}

\author{Roman~S.~Pasechnik}
\email{roman.pasechnik@fysast.uu.se} \affiliation{High Energy
Physics, Department of Physics and Astronomy, Uppsala University Box
516, SE-75120 Uppsala, Sweden}

\author{Dmitry~V.~Shirkov}
\author{Oleg~V.~Teryaev}

\affiliation{Bogoliubov Lab, JINR, Dubna 141980, Russia}

\author{Olga~P.~Solovtsova}
\author{Vyacheslav~L.~Khandramai}

\affiliation{Gomel State Technical University, Gomel 246746, Belarus}

\date{\today}

\begin{abstract}
The interplay between higher orders of the perturbative QCD (pQCD)
expansion and higher-twist contributions in the analysis of recent
Jefferson Lab data on the lowest moment of the spin-dependent proton
$\Gamma_1^{p} (Q^2)$ at $0.05<Q^2< 3\,{\rm GeV}^2$ is studied. We
demonstrate that the values of the higher-twist coefficients
$\mu^{p,n}_{2k}\,$ extracted from the data by using the
singularity-free analytic perturbation theory provide a better
convergence of the higher-twist series than with the standard
perturbative QCD. From the high-precision proton data, we extract
the value of the singlet axial charge $a_0(1\,{\rm
GeV}^2)=0.33\pm0.05$. We observe a slow $Q^2$ dependence of fitted
values of the twist coefficient $\mu_4$ and $a_0$ when going to
lower energy scales, which can be explained by the renormalization
group evolution of $\mu_4(Q^2)$ and $a_0(Q^2)$. As the main result,
a good quantitative description of all the Jefferson Lab data sets
down to $Q \simeq 350$~MeV is achieved.
\end{abstract}


\maketitle
\section{Introduction}

The spin structure of the nucleon remains the essential problem of
nonperturbative QCD and hadronic physics. One of its most
significant manifestations is the so-called spin crisis or spin
puzzle related to the surprisingly small fraction of proton
polarization carried by quarks \cite{Anselmino:1994gn,Leader08}.
This problem attracted attention to the peculiarities of the
underlying QCD description of the nucleon spin, in particular, to
the role of the gluonic anomaly (see
\cite{Anselmino:1994gn,Efremov:1989sn} and references therein). The
natural physical interpretation of these effects was the gluon
(circular) polarization, while the experimental indications of its
smallness may also point to a possible manifestation of the anomaly
via the strangeness polarization \cite{OT09}. The key point is its
consideration as a kind of heavy-quarks polarization
\cite{Polyakov:1998rb} due to the multiscale \cite{OT09} picture of
the nucleon exploring the fact that strange quark mass is much (as
the squared ratios matter) smaller than the nucleon one and, in
turn, larger than higher-twist parameters.

Higher-twist parameters (known also as the color polarizabilities)
are important ingredients of the nucleon spin structure. Their
extraction from experimental studies is relatively complicated as
they are most pronounced at low momentum transfer $Q$. Although in
this region very accurate Jefferson Lab (JLab) data are now
available, higher-twist contributions are shadowed by Landau
singularities of QCD coupling. As was shown in Ref.~\cite{Bjour},
this problem may be solved by the use of singularity-free couplings
which allowed a quite accurate extraction of higher twist (HT) and a
fairly good description of data down to rather low $Q$. The object
of investigation in \cite{Bjour} was the difference of the lowest
moments $\Gamma^{p,n}_1$ of proton and neutron structure functions
$g_1$, which corresponds to the renowned Bjorken sum rule (BSR)
\cite{Bj66}. At finite $Q^2$ the moments $\Gamma^{p,n}_1$ are
modified by higher order radiative corrections and higher-twist
power corrections, as dictated by the operator product expansion
(OPE). Such generalized ($Q^2$-dependent) BSR became a convenient
and renowned target ground for testing different possibilities of
combining both the perturbative and nonperturbative QCD
contributions in the low-energy domain (see, for example,
Refs.~\cite{Kodaira79,SofTer}).

The global higher twist analysis of the data on the spin-dependent
proton structure function $g^p_1$ at relatively large
$1<Q^2<30\,\GeV^2$, was performed in Ref.~\cite{Osipenko:2004xg}.
While the $1/Q^2$ term in the OPE works at relatively high scales
$Q^2\gtrsim 1\,\GeV^2$, higher-twist power corrections
$1/Q^4,\,1/Q^6,$ etc., start to play a significant role at lower
scales, where the influence of the ghost singularities in the
coefficient functions within the standard perturbation theory (PT)
becomes more noticeable. It affects the results of extraction of the
higher twists from the precise experimental data leading to unstable
OPE series and huge error bars \cite{Bjour}. It seems natural that
the weakening or elimination of the unphysical singularities of the
QCD coupling would allow shifting the perturbative QCD (pQCD)
frontier to a lower energy scale and getting more exact information
about the nonperturbative part of the process described by the
higher-twist series.

As was shown in Ref.~\cite{Bjour}, the situation becomes better if
one uses for a running coupling a more precise iterative solution of
the renormalisation group (RG) equation in the form of the so-called
denominator representation \cite{denom06} instead of the Particle
Data Group loop $1/L$ expansion \cite{pdg08}, especially at the
two-loop level. In this investigation, to avoid completely the
unphysical singularities at $Q=\Lambda_{QCD}\sim 400\,\MeV$ we deal
with the ghost-free analytic perturbation theory (APT)
\cite{apt96-7} (for a review on APT concepts and algorithms, see
also Ref.~\cite{Sh-revs}), which recently proved to be an intriguing
candidate for a quantitative description of light quarkonia spectra
within the Bethe-Salpeter approach \cite{BSAPT}, and the so-called
glueball-freezing model proposed recently by Yu.~A.~Simonov in
Ref.~\cite{Simonov} (below, SGF model) to avoid the renormalon
\cite{renormalon} ambiguity in QCD. Other versions of frozen
$\alpha_s$ models were developed earlier in Ref.~\cite{IR-freez}. As
it will seen below that APT and SGF approaches predict very close
couplings at $Q\gtrsim\Lambda_{QCD}$, whereas they have different
infrared-stable points at $Q=0$. Consequently, as it was shown in
Ref.~\cite{Bjour}, these models lead to very close perturbative
parts of the Bjorken sum $\Gamma_{1,pert}^{p-n}$. The higher-twist
contributions turned out to be very close, too. Here, we would like
to discuss this point in more detail.

In the current paper we study the interplay between higher orders of
the pQCD expansion and higher-twist contributions using the recent
JLab data on the lowest moments of the spin-dependent proton and
neutron structure functions $\Gamma_1^{p,n} (Q^2)$ and
$\Gamma_1^{p-n}(Q^2)$ in the range $0.05<Q^2<3\,{\rm GeV}^2$
\cite{Deur:2008rf}. Thus, we extend and generalize the analysis
started in Ref.~\cite{Bjour} by considering also the singlet channel
involving the $\Gamma_1^{p,n} (Q^2)$ for the proton (providing the
most accurate data) and the neutron structure functions separately.
This allows, in particular, determining the singlet axial charge
$a_0$ coming into both $\Gamma_1^{p,n}(Q^2)$ moments, which in the
quark-parton model is identified with the total spin carried by
quarks in the proton. For this purpose, we perform the global
analysis of the JLab precise low-energy data on $\Gamma_1^{p}(Q^2)$
\cite{JLab08data} using the advantages of the APT and SGF model, and
extract the singlet axial charge $a_0$, as well as the coefficient
$\mu^{p,n}_4$ of the $1/Q^2$ subleading twist-4 term, which contains
information on quark-gluon correlations in nucleons.

The paper is organized as follows. In Sec. 2, the lowest moments
analysis for the polarized structure functions $g^{p,n}_1$ in the
framework of the conventional PT approach is performed. In Sec. 3,
we dwell briefly on the APT, its ideas and the results of its
application to $\Gamma^{p,n}_1(Q^2)$. In Sec. 4, we apply the
formalism to the analysis of the low-energy data on the first
moments $\Gamma^{p,n}_1(Q^2)$ and compare the results with the
results of other researchers concerning the singlet axial constant
$a_0$ and gluon polarization $\Delta g$ at low $Q^2\lesssim
1\,\GeV^2$. Section 5 contains discussion and some concluding
remarks.

\section{Spin sum rules in conventional PT}

\subsection{First moments of spin structure functions $g_1^{p,n}$}

The lowest moments of spin-dependent proton and neutron structure
functions $g^{p,n}_1$ are defined as follows:
\begin{eqnarray}\label{eq1}
\Gamma_1^{p,n}(Q^2)=\int^1_0dx\, g^{p,n}_1(x,Q^2)\,,
\end{eqnarray}
with $x=Q^2/2M\nu$, the energy transfer $\nu$, and the nucleon mass
$M.$ The upper limit includes the proton/neutron elastic
contribution at $x=1$. This contribution becomes essential if the
OPE is used to study the evolution of the integral in the moderate
and low momentum transfer region $Q^2 \lesssim 1\,\GeV^2$ \cite{Ji}.
It is of special interest to analyze data with the elastic
contribution excluded, since the low-$Q^2$ behavior of ``inelastic''
contributions to their nonsinglet combination $\Gamma^{p-n}_1(Q^2)$,
i.e. BSR, is constrained by the Gerasimov-Drell-Hearn (GDH) sum rule
\cite{GDH}, and one may investigate its continuation to a low scale
\cite{SofTer}. So below we study inelastic contributions
$\Gamma^{p,n}_{inel,1}(Q^2)$ using the corresponding low-energy JLab
data \cite{JLab08data}. Note that the influence of the ``elastic''
contribution is noticeable starting from the higher-twist
$\sim\mu_6$ term which is natural due to a decrease of the elastic
contribution with growing $Q^2$ \cite{Bjour}.

At large $Q^2$ the moments $\Gamma_1^{p,n}(Q^2)$ are given
by the OPE series in powers of $1/Q^2$ with the expansion
coefficients related to nucleon matrix elements of operators of a
definite twist (defined as the dimension minus the spin of the
operator), and coefficient functions in the form of pQCD series in
$\alpha_s^n$ (see, e.g., Ref.~\cite{kataev}). In the limit $Q^2\gg
M^2$ the moments are dominated by the leading twist contribution,
$\mu_2^{p,n}(Q^2)$, which is given in terms of matrix elements of
the twist-2 axial vector current,
$\bar{\psi}\gamma^{\mu}\gamma_5\psi$. This can be decomposed into
flavor singlet and nonsinglet contributions. The total expression
for the perturbative part of $\Gamma^{p,n}_1(Q^2)$ including the HT
contributions reads
\begin{eqnarray}
\Gamma^{p,n}_{1}(Q^2)=\frac{1}{12}\left[\biggl(\pm
a_3+\frac13a_8\biggr)E_{NS}(Q^2)+\frac43
a^{inv}_0\,E_{S}(Q^2)\right]+
\sum_{i=2}^{\infty}\frac{\mu^{p,n}_{2i}(Q^2)}{Q^{2i-2}},\label{PT-Gam}
\end{eqnarray}
where $E_{S}$ and $E_{NS}$ are the singlet and nonsinglet Wilson
coefficients, respectively, calculated as series in powers of $\as$
\cite{Larin:1997qq}. These coefficient functions for $n_f=3$ active
flavors in the \msbar scheme are
\begin{eqnarray} \label{nonsinglet}
E_{NS}(Q^2)&=&1-\frac{\alpha_s}{\pi}-3.558\left(\frac{\alps}{\pi}\right)^2
-20.215\left(\frac{\alps}{\pi}\right)^3-O(\alps^4)\,, \\
E_{S}(Q^2)&=&1-\frac{\alpha_s}{\pi}-1.096\left(\frac{\alps}{\pi}\right)^2-O(\alps^3)\,.
\label{singlet}
\end{eqnarray}
The triplet and octet axial charges $a_3\equiv g_A=1.267\pm0.004$
\cite{pdg08} and $a_8=0.585\pm0.025$ \cite{Goto:1999by},
respectively, are extracted from weak decay matrix elements and
are known from $\beta$-decay measurements. As for the singlet
axial charge $a_0$, it is convenient to work with its RG invariant
definition in the \msbar scheme $a_0^{inv}=a_0(Q^2=\infty)$, in
which all the $Q^2$ dependence is factorized into the definition
of the Wilson coefficient $E_S(Q^2)$.

In contrast to the proton and neutron spin sum rules (SSRs), the
singlet and octet contributions are canceled out, giving rise to
more fundamental BSR
\begin{eqnarray}
\Gamma^{p-n}_{1}(Q^2)=\frac{g_A}{6}E_{NS}(Q^2)+
\sum_{i=2}^{\infty}\frac{\mu^{p-n}_{2i}(Q^2)}{Q^{2i-2}},\label{BSR}
\end{eqnarray}
which is analyzed here along with the proton SSR in more detail than
in Ref.~\cite{Bjour}. The first nonleading twist term \cite{Shuryak}
can be expressed \cite{chen06}
\begin{eqnarray}\nonumber
 \mu_4^{p-n}\approx\frac{4\,M^2}{9}f_2^{p-n},
\end{eqnarray}
in terms of the color polarizability $f_2$.

The RG $Q^2$ evolution of the axial singlet charge $a_0(Q^2)$ and
nonsinglet higher-twist $\mu^{p-n}_4(Q^2)$ is \cite{Shuryak}
\begin{eqnarray}\label{a0-evol}
a_0(Q^2)&=&a_0(Q_0^2)\exp\left\{\frac{\gamma_2}{(4\pi)^2\beta_0}[\alps(Q^2)-\alps(Q_0^2)]\right\},
\quad\gamma_2=16n_f,\\ [0.2cm]
\label{mu4pn-evol}
\mu_4^{p-n}(Q^2)&=&\mu_4^{p-n}(Q_0^2)\left[\frac{\alps(Q^2)}{\alps(Q_0^2)}\right]^{\gamma_0/8\pi\beta_0},\quad
\beta_0=\frac{33-2n_f}{12\pi},\quad
\gamma_0=\frac{16}{3}C_F\,.
\end{eqnarray}
In the NLO we may write
\begin{eqnarray}
a_0(Q^2)&\simeq& a_0(Q_0^2)\left[1+\Delta_1(Q^2)+{\cal
O}(\alps^2)\right],\\
\Delta_1(Q^2)&=&\frac{\gamma_2}{(4\pi)^2\beta_0}[\alps(Q^2)-\alps(Q_0^2)],\quad
\frac{\gamma_2}{(4\pi)^2\beta_0}=\frac{4}{3\pi}. \nonumber
\label{a0-evol-1L}
\end{eqnarray}

As a first step of our analysis, in Eq.~(\ref{PT-Gam}) we will
neglect the weak dependence of $\mu^{p,n}_{2i}$ on $\log Q^2$. Note
that the evolution of the higher-twist terms $\mu_{6,8,\,...}$ in
Eq.~(\ref{PT-Gam}) is still unknown. As a next step we discuss the
possible influence of the $\mu_4(Q^2)$ evolution on our results. The
$Q^2$ evolution of the proton higher-twist term $\mu^p_4(Q^2)$ is
assumed to be the same as the evolution of the nonsinglet twist
$\mu^{p-n}_4(Q^2)$. This may be justified by the relative smallness
of the singlet higher-twist term.

\begin{table}[h!]
\caption{\small\sf Current NLO fit results for the axial singlet
charge $a_0$.}
\begin{center}\label{table1}
\begin{tabular}{|c|c|c|c|c|c|} \hline
 Reference & $\;$LSS \cite{LSS06}$\;$ & $\;$DSSV \cite{DSSV}$\;$ & $\;$AAC \cite{AAC08}$\;$ &
 $\;$HERMES  \cite{HERMES06}$\;$ & $\;$COMPASS  \cite{COMPASS06}$\;$\\ \hline
 $Q_0^2,\,\GeV^2$ & 1.0 & 10.0 & 4.0 & 5.0 & 3.0 \\ \hline
 $a_0$ & $0.24\pm0.07$ &0.24 & $0.25\pm0.05$ &  $0.32\pm0.04$ & $0.35\pm0.06$ \\ \hline
\end{tabular} \end{center}
\end{table}

Let us discuss current results for the nucleon spin structure and
higher twists. In Table~\ref{table1}, we list the fit results for
the axial singlet charge $a_0$ from the literature including all
global NLO PT analyses and the recent results obtained directly from
deuteron data on $\Gamma_1^d$ by COMPASS \cite{COMPASS06} and HERMES
\cite{HERMES06}. The global fit results for $a_0$ are somewhat lower
than that from the deuteron data. It was mentioned in the most
recent review \cite{Leader08} that the reason for such a discrepancy
is not completely understood. Further, we analyze this issue in more
detail.

\begin{table}[h!]
\caption{\small\sf Current NLO fit results for the highest-twist
term $\mu_4/M^2$. The uncertainties are statistical only.}
\begin{center}\label{table2}
\begin{tabular}{|c|c|c|c|c|c|} \hline
 Target &$\quad$Proton \cite{proton}$\quad$ &
 $\quad$Neutron \cite{neutron}$\quad$ & $\quad$p -- n \cite{Bj04-tw}$\quad$ &
 $\quad$p -- n \cite{Bj08-tw}$\quad$ & $\quad$p -- n \cite{Bjour}$\quad$ \\ \hline
 $Q^2,\,\GeV^2$& 0.6 -- 10.0 & 0.5 -- 10.0 & 0.5 -- 10.0 & 0.66 --
 10.0 & 0.12 -- 3.0
  \\ \hline
 $\mu_4/M^2$ & $-0.065\pm0.012$ & $0.019\pm0.002$ & $-0.06\pm0.02$ &  $-0.04\pm0.01$ & ~$-0.048\pm0.002$~  \\ \hline
\end{tabular} \end{center}
\end{table}

A detailed higher-twist analysis based on the combined SLAC and JLab
data [on proton, neutron $\Gamma_1^{p,n}(Q^2)$ \cite{JLab-old-data}
and nonsinglet $\Gamma_1^{p-n}(Q^2)$ moments \cite{Bj08-tw}] was
performed in Refs.~\cite{proton, neutron,Bj04-tw,Bj08-tw}. In
Table~\ref{table2}, we show the current results for the twist-4
coefficient $\mu_4/M^2$ at $Q^2=1\,\GeV^2$ extracted from
$\Gamma_1^{p,n}$ data. As we have seen from our previous analysis
\cite{Bjour}, a satisfactory description of the low-energy JLab data
on the Bjorken sum rule down to $Q_{min}\sim\Lambda_{QCD}\simeq
350\,\MeV\,$ can be achieved by using APT and taking into account
only three higher-twist terms $\mu^{p-n}_{4,6,8}$. Including only
the twist-4 term $\mu^{p-n}_4/M^2$, this method allowed us to get
its value with noticeably higher accuracy than in the standard PT
approach, shifting the applicability of the pQCD expansion down to
$Q^2_{min}=0.47\,\GeV^2$. The higher-twist analysis of the most
recent precise JLab experimental data on the proton spin sum rule
\cite{JLab08data} has not been carried out yet in the literature.
This gives us a reasonable motivation for a detailed data analysis
and studying the higher-twist effects at low-energy scale both in
the standard PT, APT and ``infrared-frozen'' $\alpha_s$  approaches.

\subsection{The running coupling}

The infrared behavior of the strong coupling is crucial for the
extraction of the nonperturbative information from the low-energy
data. Within the pQCD, the $\as$ coupling can be found by a solution
of the RG equation
\begin{eqnarray}\nonumber
\frac{d\alpha_s}{dL}=-\beta_0\alpha_s^2(1+b_1\alpha_s+b_2\alpha_s^2+\,...)\,,
\end{eqnarray}
where $L=\ln(Q^2/\Lambda^2)$ and $b_k=\beta_k/\beta_0$. The standard
PT running coupling \as is usually taken in the form [see, for
example, Eq.~(6) in the recent review \cite{bethke09} or Eq.~(9.5)
in the PDG review \cite{pdg08}] expanded in a series over $\ln
L/L\,$, i.e.
\begin{eqnarray}\label{9.5pdg}
\as^{(3)}(L)=\frac{1}{\beta_0L}-\frac{b_1}{\beta_0^2}\frac{\ln
L}{L^2}+\frac{1}{\beta_0^3L^3}\left[b_1^2(\ln^2L-\ln
L-1)+b_2\right].
\end{eqnarray}
Here, the $1/L^2$ term corresponds to the 2-loop contribution and
the $1/L^3$ term is usually referred to as ``the 3-loop one.''
Actually, the pieces of genuine 2-loop contribution proportional to
$b_1\,$ are entangled with the higher-loop ones. This defect is
absent in the more compact denominator representation
\cite{denom06}, which at 2,~3-loop levels has the following forms:
\begin{eqnarray}\nonumber
&&\frac{1}{\as^{(2),D}(L)}=\beta_0\,L+b_1\ln\left(L+\frac{b_1}{\beta_0}\right),
~~\frac{1}{\as^{(3),D}(L)}=\beta_0\,L+b_1\ln\left(L+\frac{b_1}{\beta_0}\,\ln
L\right)+\frac{b_1^2-b_2}{\beta_0\,L},\\
\label{Denom}
\end{eqnarray}
which, being generic for the PDG expression (\ref{9.5pdg}), are
closer to the corresponding iterative RG solutions and, hence, more
precise. Advantages of formulas (\ref{Denom}) in the higher-twist
analysis of the Bjorken sum rule were demonstrated in our previous
work \cite{Bjour}.

\begin{figure}[!ht]
 \centerline{\epsfig{file=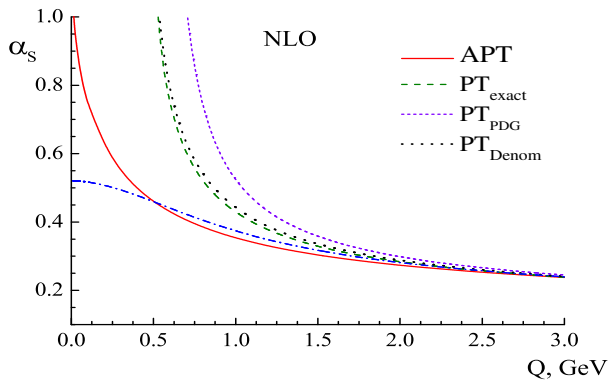,height=5cm,width=8cm}}
 \caption{\footnotesize The NLO running coupling $\as$ in different
 approaches.}
 \label{fig:as-2L}
\end{figure}

In Fig.~\ref{fig:as-2L}, we compare the behavior of the two-loop
running coupling $\as$ at low $Q^2$ scales in different approaches.
The long-dashed line is the exact two-loop PT result, the dotted
line is the denominator representation (\ref{Denom}) (referred to as
``Denom'' below), and the short-dashed line is the PDG expression
(\ref{9.5pdg}). As one can see from this figure, the NLO Denom
coupling is much closer to the corresponding numerical RG solution
than the $1/L$-expanded PDG expression.

In Fig.~\ref{fig:as-2L}, we also show two models of the
infrared-stable running coupling. One of them is the Simonov
``glueball-freezing model'' (SGF-model) \cite{Simonov}, represented
by the dash-dotted line, with the $1/L$-type loop expansion for the
``infrared-frozen'' coupling similar to PDG
\begin{eqnarray} 
 \phantom{AAAAAA}\alpha_B(Q^2)=\alpha_s^{(2)}(\bar{L})\,,\quad
 \bar{L}=\ln\left(\frac{Q^2+M_0^2}{\Lambda^2}\right),
 \label{simon}
\end{eqnarray}
where the two-loop $\alpha_s^{(2)}$ is taken in the form of the
first two terms in Eq.~(\ref{9.5pdg}) with logarithm modified by a
``glueball mass'' $M_0\sim 1\,\GeV$. Note, the usual PT expansion in
powers of $\alpha_B$ in the coefficient functions (\ref{nonsinglet})
and (\ref{singlet}) is adopted. The solid line corresponds to the
second model of the infrared-stable coupling -- the APT running
coupling, which will be discussed in detail below in the next
section.

As one can see from Fig.~\ref{fig:as-2L}, the SGF and APT couplings
are very similar in the low-energy domain $\Lambda_{QCD}<Q\lesssim
1$ GeV though their infrared limits are different. Also, a
comparison of APT and PT couplings over a wide range of $Q^2$,
$1\leq Q^2 \leq 10^4$ GeV$^2$, can be found in Ref.~\cite{APT-GLS}.

Note, we extract values of $\Lambda_{QCD}$ corresponding to
different models of the running coupling, by evolution from the
world experimental data on $\alpha_s(M_Z^2)$ as a normalization
point in each particular order of PT.

\subsection{Stability and duality}

\begin{figure}[!h]    %
\includegraphics[width=0.495\textwidth]{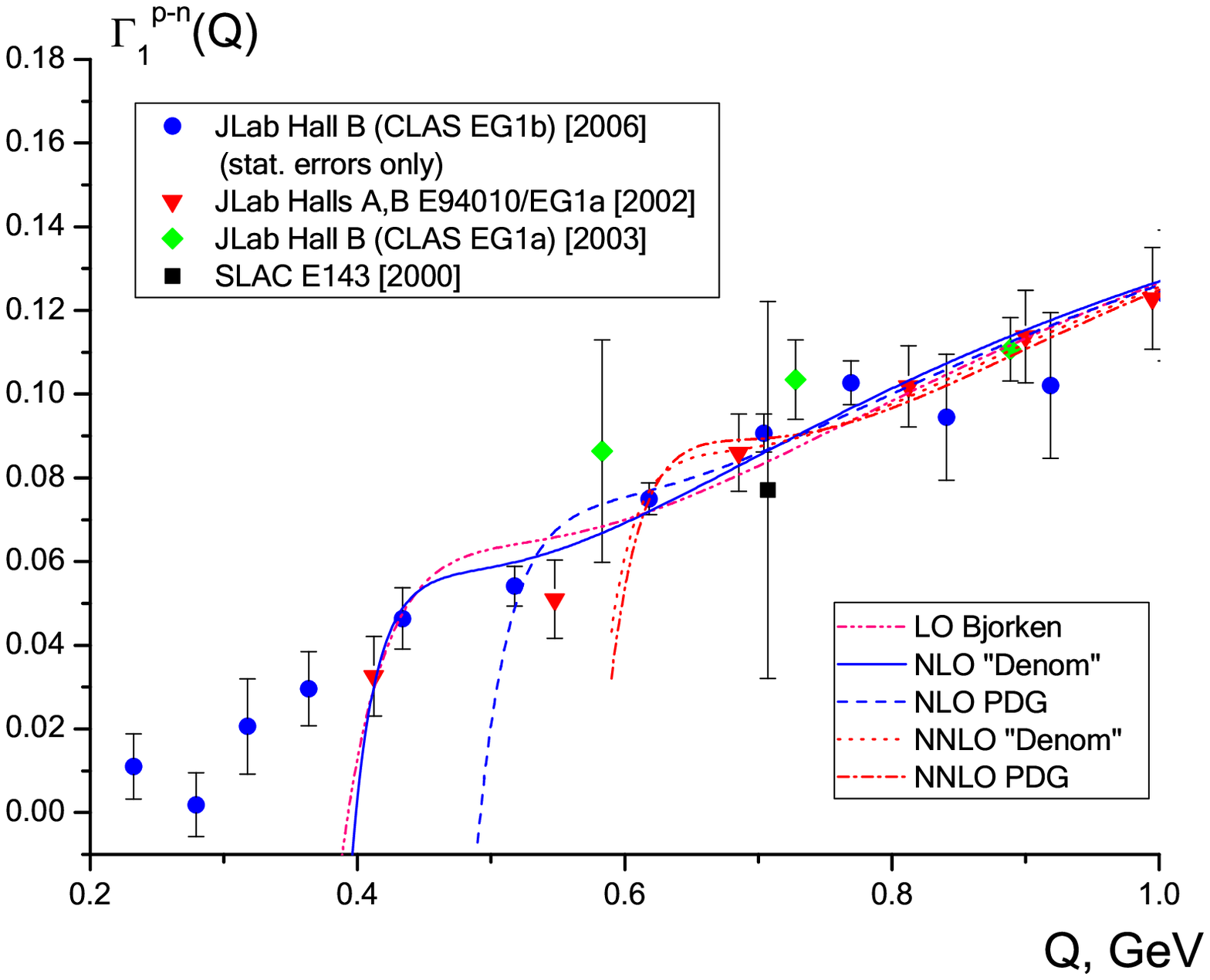}
\includegraphics[width=0.495\textwidth]{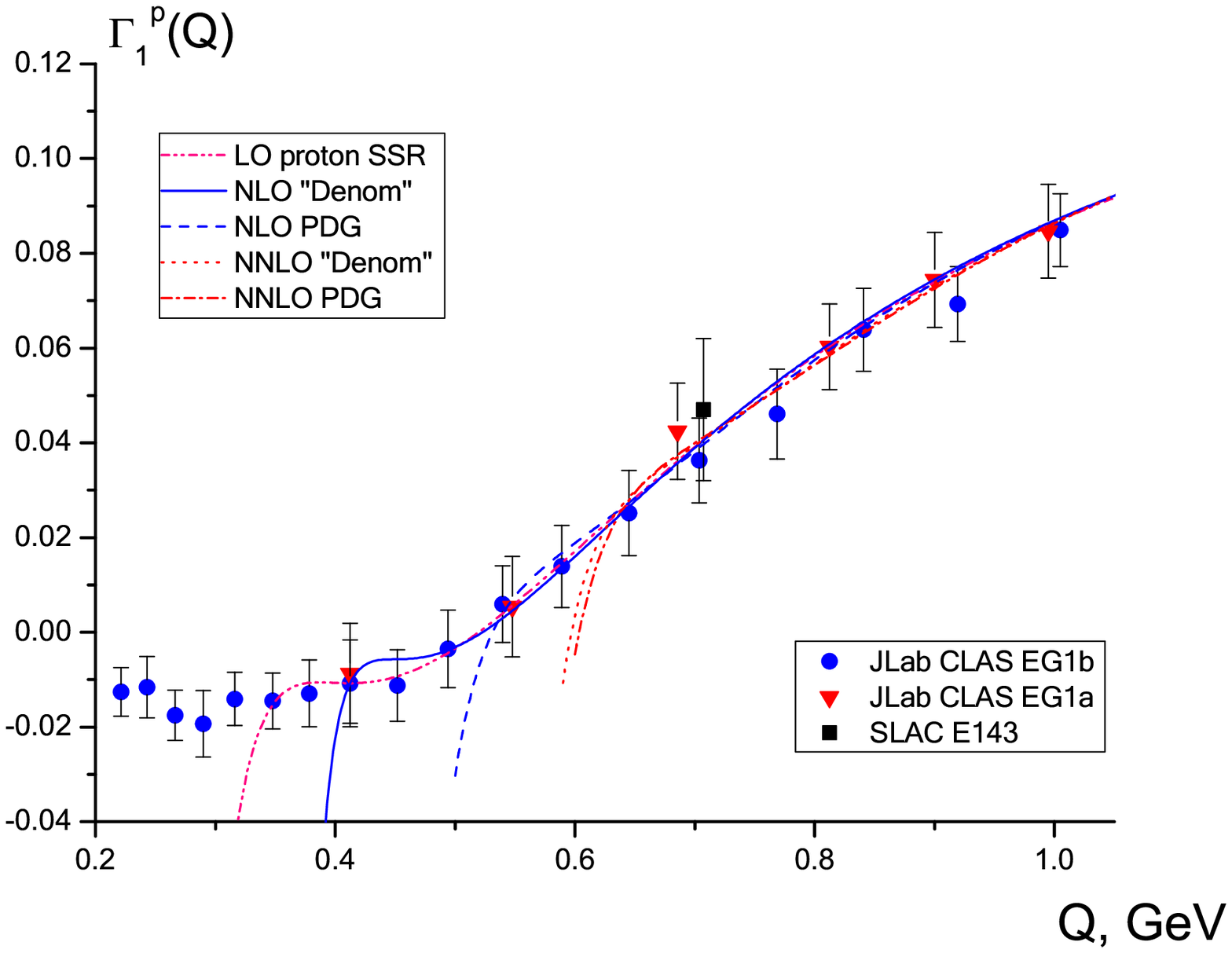}
   \caption{\label{fig:loop-sen}
  \footnotesize Best fits of JLab and SLAC data
 on BSR $\Gamma_1^{p-n}(Q^2)$ (left panel) and proton SSR
 $\Gamma_1^p(Q^2)$ (right panel) calculated at various loop orders.}
\end{figure}

In the following, when calculating the observables in any particular
order of perturbation theory, we will employ the prescription for
the coefficient functions in the infrared region, where the order of
the power $\alps$ series in the coefficient functions is matched
with the loop order in $\alps$ itself. For example, for the
nonsinglet coefficient function in the Bjorken sum rule, we write
consequently (for details, see Ref.~\cite{HERMES06})
\begin{eqnarray}
&&E^{LO}_{NS}=1,\quad
E^{NLO}_{NS}=1-\frac{\alpha^{NLO}_s}{\pi},\quad
E^{N^2LO}_{NS}=1-\frac{\alpha^{N^2LO}_s}
{\pi}-3.558\Big(\frac{\alps^{N^2LO}}{\pi}\Big)^2,\,
\ldots ~~~
\label{our-presc}
\end{eqnarray}
We see that the leading singular behavior in the coefficient
function $\sim\ln^n L/L^m$ when $L\to0$ comes from the highest
power of $\alps$. So in the infrared domain the influence of
singularities gets stronger in higher orders of perturbation
theory that may affect the data analysis below $1\,\GeV^2$. This
fact explains our observation made in Ref.~\cite{Bjour}, where we
showed that the higher PT orders yield a worse description of the
BSR data in comparison with the leading order. We observe a
similar picture for the precise JLab data on $\Gamma_1^p(Q^2)$
\cite{JLab08data} probably implying the asymptotic character of
the series in powers of $\as$ (see Fig.~\ref{fig:loop-sen}).

\begin{table}[h!]
\begin{center}
\caption{\small\sf Dependence of the best fit results of BSR
$\Gamma^{p-n}_1(Q^2)$ and proton SSR $\Gamma^{p}_1(Q^2)$ data
(elastic contribution excluded) on the order of perturbation theory
[NLO and NNLO Denom couplings (\ref{Denom}) are used]. The
corresponding fit curves are shown in Fig.~\ref{fig:loop-sen}. The
minimal borders of fitting domains in $Q^2$ are settled from the
{\it ad hoc} restriction $\chi^2\leqslant1$ and monotonous behavior
of the resulting fitted curves.} \label{tab:loops} \vspace{5mm}
\begin{tabular}{|c||c|c|c|c|c|c|}
 \hline Target &$\;$Method$\;$& $\; Q^2_{min},\,\GeV^2\;$ & $\quad a_0^{inv}\quad$
 & $\quad\mu_4/M^2\quad$ & $\quad\mu_6/M^4\quad$ & $\quad\mu_8/M^6\quad$
 \tabularnewline\hline \hline
        & LO              & 0.121 & $0.29(2)$ & $-0.089(3)$ & $0.016(1)$ & $-0.0010(1)$
 \tabularnewline
 proton & NLO             & 0.17  & $0.38(2)$ & $-0.070(5)$ & $0.010(2)$ & $0.0004(3)$
 \tabularnewline
        & NNLO            & 0.38  & $0.37(5)$ & $-0.034(19)$ & $-0.025(20)$ & $0.017(6)$
 \tabularnewline \hline \hline
        & LO              & 0.17   &  --    &  $-0.126(5)$  &  $0.037(3)$  & $-0.004(1)$
 \tabularnewline
 p -- n & NLO             & 0.17   &  --    &  $-0.076(5)$  &  $0.019(3)$  & $-0.001(1)$
 \tabularnewline
        & NNLO            & 0.38   &  --    &  $-0.026(11)$ &  $-0.035(15)$& $0.026(5)$
 \tabularnewline\hline
\end{tabular}
\end{center}
\end{table}

The corresponding fit results for HT terms, extracted in different
orders of PT, are listed in Table~\ref{tab:loops}. We see that with
raising the loop order the values of $\mu^p_{4,8}$ terms increase,
whereas $\mu^p_6$ decreases, yielding a ``swap'' between the higher
orders of PT and HT terms. Such a ``swap'' between PT and HT terms
(decreasing HT term by including more terms of PT and using
resummation of PT series) was previously observed in
Refs.~\cite{Kotikov:1992ht,Parente:1994bf}. A similar situation
holds when fitting $\Gamma^p_1(Q^2)$ data over the fixed range
$0.8\,\GeV<Q<2.0\,\GeV\,$, where it is sufficient to take into
account only one twist term $\mu_4$.
\begin{figure}[!h]    %
\includegraphics[width=0.495\textwidth]{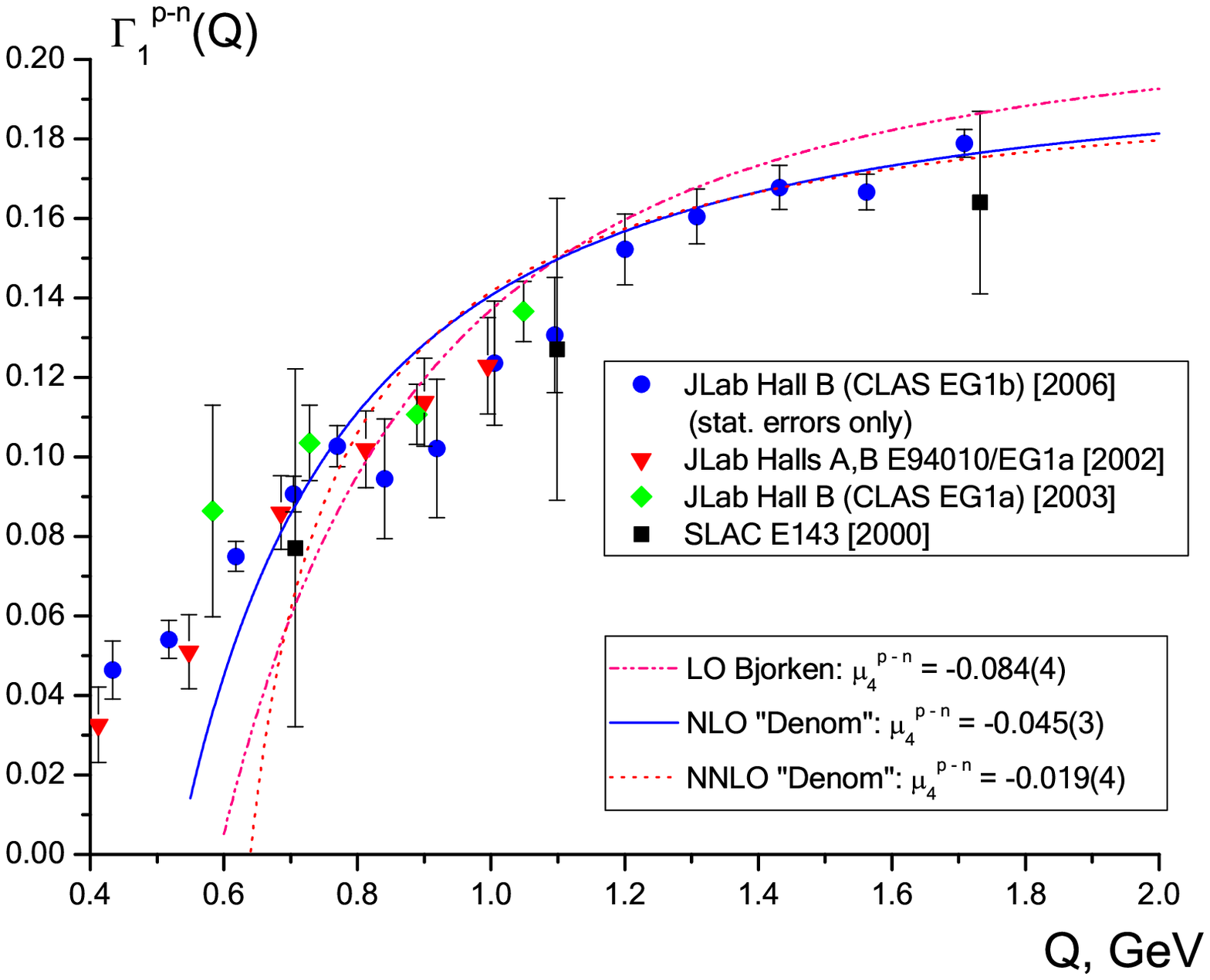}
\includegraphics[width=0.495\textwidth]{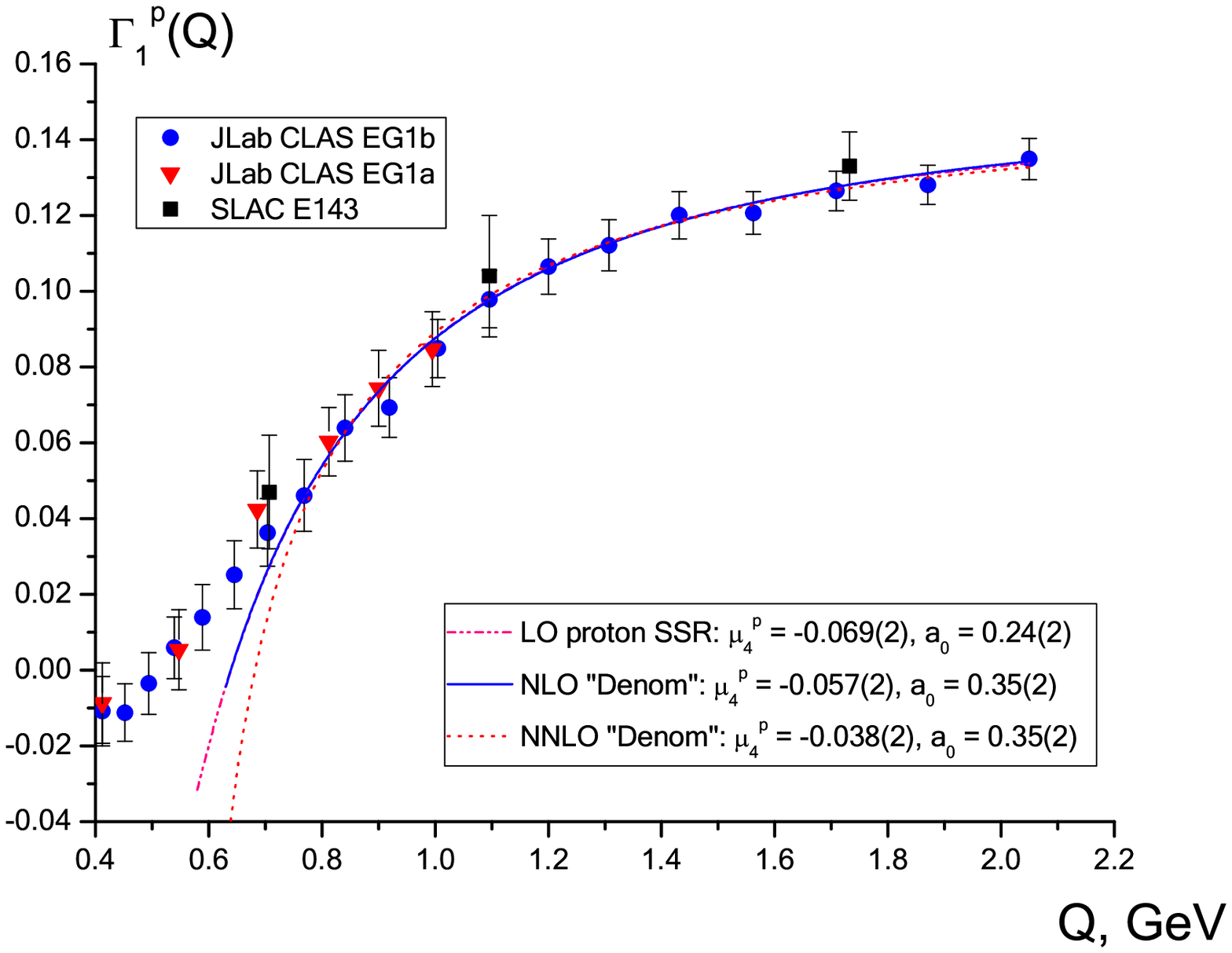}
   \caption{\label{fig:dual}
  \footnotesize Best fits of JLab and SLAC data
 on BSR $\Gamma_1^{p-n}(Q^2)$ (left panel) and proton SSR
 $\Gamma_1^p(Q^2)$ (right panel) calculated in various loop orders
 with fixed $Q_{min}=0.8$ GeV.}
\end{figure}

In Fig.~\ref{fig:dual}, we show fits of BSR data (left panel) and
proton SSR data (right panel) in different orders of perturbation
theory taking only into account the $\mu_4$ term. One can see there
that the higher-loop contributions are effectively ``absorbed'' into
the value of $\mu_4$ which decreases in magnitude with increasing
loop order while all the fitting curves are very close to each
other. This observation reveals a kind of ``duality'' between the
perturbative $\as$ series and nonperturbative $1/Q^2$ series. A
similar phenomenon was observed before for the structure function
$F_3$ in Refs.~\cite{Kataev:1997nc,SidKat}.

This also means the appearance of a new aspect of quark hadron duality,
the latter being the necessary ingredient of all the QCD applications
in the low-energy domain. Usually, it is assumed \cite{Shifman:1978bx}
that the perturbative effects are less important there than the power
ones due to a nontrivial structure in the QCD vacuum.

 In our case, the PT corrections essentially enter into the
 game, so that the pQCD higher order terms are relevant in the domain where the
 concepts of  traditional hadronic physics are usually applied.

\begin{table}[!h]
\begin{center}
\begin{minipage}[t]{16.5 cm}
\caption{\small\sf Dependence of the best $(3+1)$-parametric fit
results of $\Gamma^{p}_{1}(Q^2)$ data (elastic contribution
excluded) on $\Lambda_{n_f=3}$ in NLO Denom PT.} \label{tab:Lam}
\vspace{5mm}
\end{minipage}
\begin{tabular}{|c|c|c|c|c|c|}
 \hline$\;\Lambda_{QCD},\,\MeV\;$& $\quad Q^2_{min},\,\GeV^2\quad$ & $\quad a_0^{inv}\quad$
 & $\quad\mu_4/M^2\quad$ & $\quad\mu_6/M^4\quad$ & $\quad\mu_8/M^6\quad$
 \tabularnewline \hline \hline
  300 &  0.14   & $0.40(2)$ & $-0.077(3)$ & $0.014(1)$ & $-0.0005(2)$
 \tabularnewline
  400 &  0.24   & $0.39(3)$ & $-0.064(8)$ & $0.006(5)$ & $0.002(1)$
 \tabularnewline
  500 &  0.35   & $0.34(4)$ & $-0.028(13)$ & $-0.033(11)$ & $0.019(3)$
 \tabularnewline \hline
\end{tabular}
\end{center}
\end{table}

The interplay between partonic and hadronic degrees of freedom in the
description of GDH SR and BSR may also be observed in the surprising
similarity between the results of ``resonance'' \cite{Burkert:1992tg}
and ``parton'' \cite{SofTer} approaches.

One may ask to what extent these results are affected by the
unphysical singularities when approaching $Q\sim \Lambda_{QCD}$ in
the PT series for $\Gamma^{p,n}_{1,PT}$. Their influence becomes
essential at $Q<1\,\GeV$ where the HT terms start to play an
important role. The minimal border of the fitting domain $Q_{min}$
is tightly connected with the value of $\Lambda_{QCD}$; i.e. it is a
scale, below which the influence of the ghost singularities becomes
too strong and destroys the fit. To see how the $Q^2_{min}$ scale
and fit results for the $\mu$ terms change with varying
$\Lambda_{QCD}$, we have performed three different NLO fits with
$\Lambda_{QCD}=300,\,400,\,500\,\MeV$ (see Table~\ref{tab:Lam}). It
turns out that the term $\mu_4$ is quite sensitive to the Landau
singularity position, and its value noticeably increases with
increasing $\Lambda_{QCD}$. The APT and ``soft-frozen'' models are
free of such a problem, thus providing a reliable tool of
investigating the behavior of HT terms extracted directly from the
low-energy data \cite{Bjour}. This provides a motivation for the
analysis performed in the next section.

\section{Moments $\Gamma_1^{p,n}(Q^2)$ in Analytic Perturbation Theory}

The moments of the structure functions are analytic functions in the
complex $Q^2$ plane with a cut along the negative real axis, as was
demonstrated in Ref.~\cite{W78} (see also Ref.~\cite{Ashok_suri}).
On the other hand, the standard PT approach does not support these
analytic properties. The influence of requiring these properties to
hold in the DIS description was studied previously by Igor Solovtsov
and coauthors in Refs.~\cite{APT-GLS,MSS}. Here we continue this
investigation by applying the APT method, which gives the
possibility of combining the RG resummation with correct analytic
properties of the QCD corrections, to the low-energy data on nucleon
spin sum rules $\Gamma_1^{p,n}(Q^2)$.

In the framework of the analytic approach we can write the
expression for $\Gamma_1^{p,n}(Q^2)$ in the form
\begin{eqnarray}
\Gamma^{p,n}_{1,APT}(Q^2)=\frac{1}{12}\left[\biggl(\pm
a_3+\frac13a_8\biggr)E^{APT}_{NS}(Q^2)+\frac43
a^{inv}_0\,E^{APT}_{S}(Q^2)\right]+
\sum_{i=2}^{\infty}\frac{\mu^{APT;\,p,n}_{2i}(Q^2)}{Q^{2i-2}} \, ,
\label{APT-Gam}
\end{eqnarray}
which is analogous to one in the standard PT (\ref{PT-Gam}). The
corresponding NNLO APT modification of the singlet and nonsinglet
coefficient functions is
\begin{eqnarray}
E^{APT}_{NS}(Q^2)&=&1-0.318\,{\cal A}^{(3)}_1(Q^2)-0.361\,{\cal
A}^{(3)}_2(Q^2)-\,...\,, \\
E^{APT}_{S}(Q^2)&=&1-0.318\,{\cal A}^{(3)}_1(Q^2)-0.111\,{\cal
A}^{(3)}_2(Q^2)-\,...\, , \label{E-APT}
\end{eqnarray}
where ${\cal A}^{(3)}_k$ is the analyticized $k$th power of 3-loop
PT coupling in the Euclidean domain
 \begin{eqnarray}
  \label{Akn}
  \ac^{(n)}_k(Q^2)=\frac{1}{\pi} \int^{+\infty}_0
  \frac{\mathrm{Im}([\as^{(n)}(-\sigma,n_f)]^k)\,d\sigma}{\sigma+Q^2},\qquad
  n=3\,.
 \end{eqnarray}
In the one-loop case, the APT Euclidean functions are simple enough
\cite{apt96-7}:
\begin{eqnarray}
 &&\acal_1^{(1)}(Q^2)=\frac{1}{\beta_0}\left[\frac{1}{L}+
 \frac{\Lambda^2}{\Lambda^2-Q^2}\right]\,,\quad
 L=\ln\left(\frac{Q^2}{\Lambda^2}\right),\label{AE1-2}\\
 &&\acal_2^{(1)}(l)=\frac{1}{\beta_0^2}\left[\frac{1}{L^2}-
 \frac{Q^2\,\Lambda^2}{(Q^2-\Lambda^2)^2}\right],\;
 \acal_{k+1}^{(1)}=-\,\frac{1}{k\,\beta_0}\,
 \frac{d\,\acal_k^{(1)}} {d L} \, ,\nonumber
\end{eqnarray}
i.e. the higher functions ${\cal A}_k$ are related to the lower ones
recursively by differentiating. Analogous two- and three-loop level
expressions involve the special Lambert function and are more
intricate, and they can be found in Refs.~\cite{Magr:00,K-Magr:01}.
It should be stressed that the APT couplings are stable with respect
to different loop orders at low-energy scales $Q^2\lesssim
1\,\GeV^2$ \cite{Sh-revs}. This feature is absent in the standard PT
approach, as reflected in Fig.~\ref{fig:loop-sen}.

Meanwhile, even for the three-loop APT case, there exists a
possibility to employ {\it the effective log approach} proposed by
Igor Solovtsov and one of the authors in Ref.~\cite{SolSh99}. In the
present context, in the region $\,Q<5\,\,\GeV$ one may use simple
model one-loop expressions (\ref{AE1-2}) with some {\it effective
logarithm} $L^*\,$:
\begin{eqnarray} \label{model}                       
 \acal_{1,2,3}^{(3)}(L)\to\acal_{1,2,3}^{mod}=\,\acal_{1,2,3}^{(1)}(L^*)
 \,,\quad L^*\simeq 2\,\ln(Q/\Lambda^{(1)}_{eff}),\quad\Lambda^{(1)}_{eff}\simeq
 0.50\,\Lambda^{(3)}.
\end{eqnarray}
Thus, instead of the exact three-loop expressions for the APT
functions, in Eq.~(\ref{E-APT}) one can use the one-loop expressions
(\ref{AE1-2}) with the effective $\Lambda$ parameter
$\Lambda_{mod}=\Lambda^{(1)}_{eff}\,$ whose value is given by the
last relation (\ref{model}). This model was successfully applied for
higher-twist analysis of low-energy data on BSR in our previous work
\cite{Bjour}, and also in the $\Upsilon$ decay analysis in
Ref.~\cite{ShZ05}.

The maximal errors of the model (\ref{model}) for the first and
the second functions are $\delta\ac^{mod}_1/\ac^{mod}_1\simeq4\%$
and $\delta\ac^{mod}_2/\ac^{mod}_2\simeq8\%$ at
$Q\sim\Lambda_{n_f=3}\,,$ which seem to be sufficiently
accurate. Indeed, as far as $\acal_1(Q=400\,\MeV)=0.532\,$ and
$\acal_2(400\,\MeV)=0.118\,,$ the total error in $\Gamma^p_{\rm
1,APT}\,$ is mainly determined by the first term, being of the
order $\delta\Gamma_1^p/\Gamma_1^p
\simeq\delta\ac^{mod}_1/\pi\sim1\,\%\,,$ i.e., less than the data
uncertainty.
\begin{figure}[!h]\leavevmode\begin{center}
         \begin{minipage}[b]{0.49\textwidth}
                \phantom{}\hspace{-0.5cm}%
\centering\includegraphics[width=0.96\textwidth]{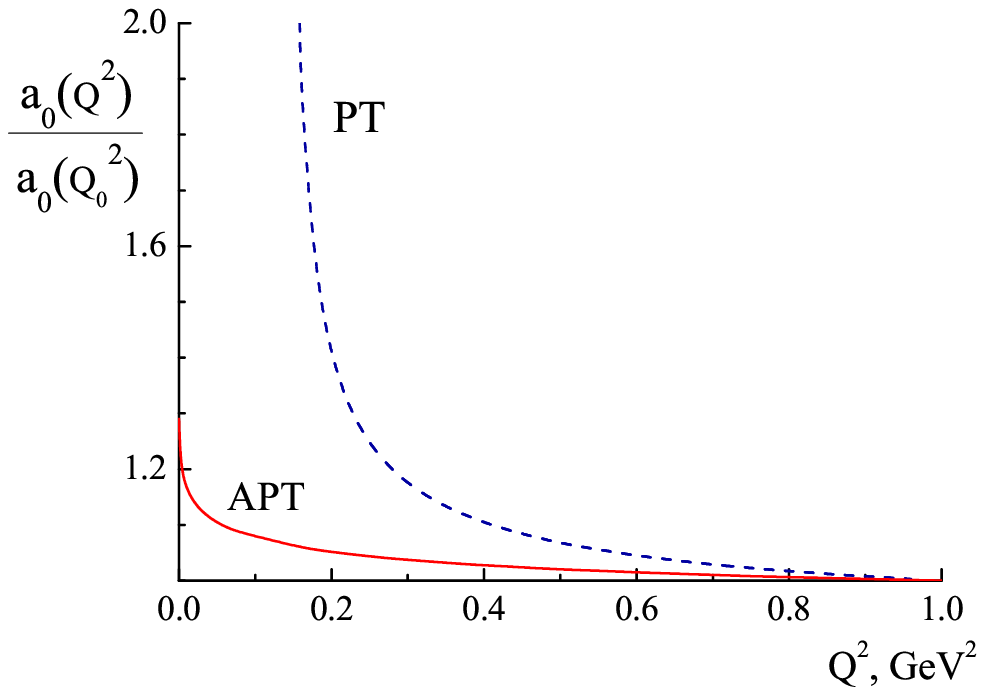}
         \end{minipage}%
     \begin{minipage}[b]{0.49\textwidth}
     \vspace*{-0.0cm}
\centering\includegraphics[width=0.96\textwidth]{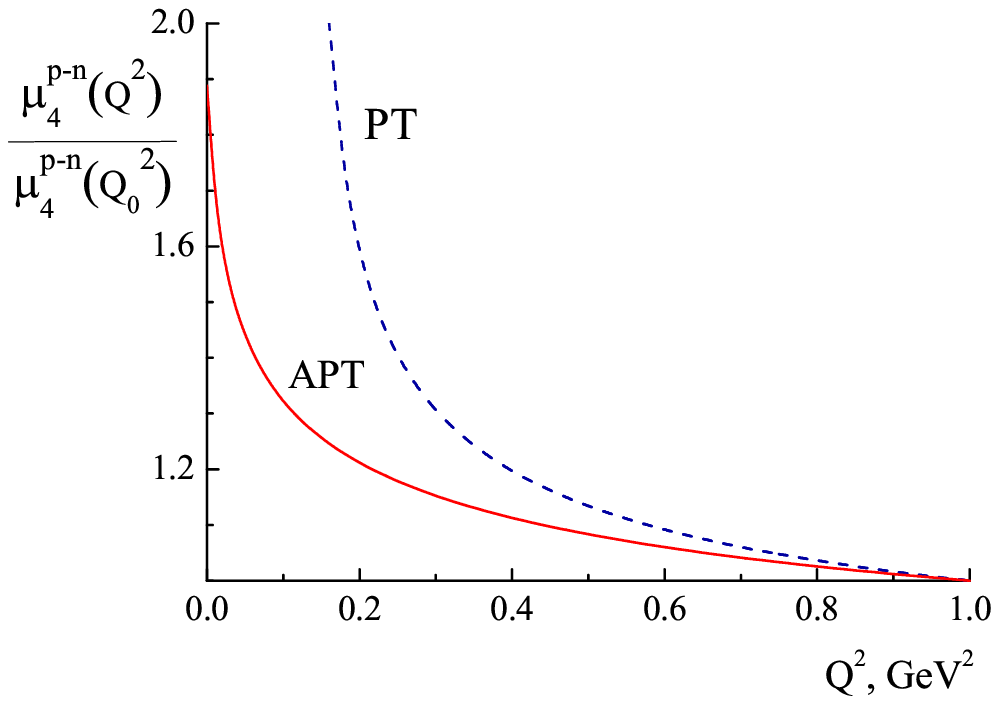} 
    \end{minipage}
             \begin{minipage}[t]{0.45\textwidth}
 \caption{\footnotesize Evolution of $a_0(Q^2)$
 normalized at $Q_0^2=1\,\GeV^2$.} \label{fig:evol-a0}
        \end{minipage}%
\phantom{}\hspace{0.5cm}%
     \begin{minipage}[t]{0.45\textwidth}
 \caption{\footnotesize Evolution of $\mu^{p-n}_4(Q^2)$
 normalized at $Q_0^2=1\,\GeV^2$.} \label{fig:evol-mu4}
    \end{minipage}      \end{center}
    \end{figure}

In order to take into account the one-loop $Q^2$ evolution of the
axial singlet charge $a_0(Q^2)$, we use expression
(\ref{a0-evol-1L}) substituting the one-loop analytic coupling
$\acal_{1}^{(1)}(L)$. The contribution of the $\sim\acal_1$ term to
$a_0(Q^2)$ at, for example, $Q^2=0.1\,\GeV^2$ with normalization
point at $Q_0^2=1\,\GeV^2$ is $\Delta_1(0.1\,\GeV^2)\simeq 0.11$;
i.e. the evolution contributes about 10\% when one shifts the pQCD
border down to $\Lambda_{QCD}$ (see Fig.~\ref{fig:evol-a0}).

For the evolution of the twist-4 term $\mu_4(Q^2)$
(\ref{mu4pn-evol}), we have to ``analyticize'' the fractional power
$(\alps)^{\nu}$. For this purpose we apply the fractional APT
approach developed in Ref.~\cite{Bakulev}. At the one-loop level in
the Euclidean domain we have
\begin{eqnarray}
\acal_{\nu}^{(1)}(L)=\frac{1}{L^{\nu}}-\frac{F(e^{-L},1-\nu)}{\Gamma(\nu)}.
\label{fAPT}
\end{eqnarray}
Here $F(z,\nu)$ is the Lerch transcendent function. In this case,
the evolution of the nonsinglet twist-4 term in BSR reads
\begin{eqnarray}
\mu_{4,APT}^{p-n}(Q^2)= \mu_{4,APT}^{p-n}(Q_0^2)\,
\frac{\acal_{\nu}^{(1)}(Q^2)}{\acal_{\nu}^{(1)}(Q_0^2)},\qquad\nu=\frac{32}{81}.
\label{APT-evol-mu}
\end{eqnarray}
The corresponding evolution is shown in Fig.~\ref{fig:evol-mu4}.
As follows from this figure, the evolution from $1\,\GeV$ to
$\Lambda_{QCD}$ increases the absolute value of
$\mu_{4,APT}^{p-n}$ by about 20~\%.

\section{Numerical results}

\subsection{Nonsinglet case: the Bjorken sum rule}
\begin{figure}[!h]
 \centerline{\epsfig{file=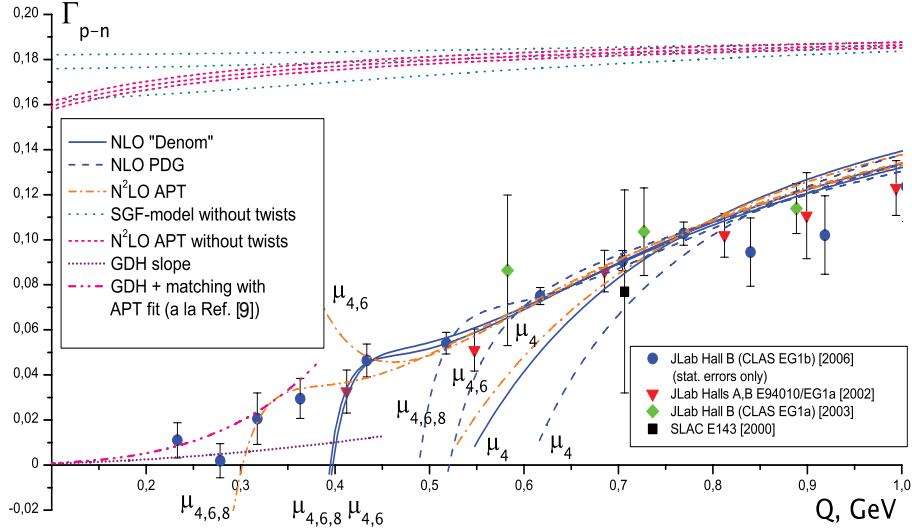,height=7cm,width=12cm}}
 \caption{\footnotesize Best 1,2,3-parametric fits of the JLab and SLAC
 data on Bjorken SR calculated with different models of running coupling.}
 \label{fig:Bjtot}
\end{figure}

\begin{table}[!h]
\caption{\small\sf Combined fit results of BSR for the HT terms in
APT, the SGF model and the standard PT approach.}
\begin{center}\label{tab:Bjtot}
\begin{tabular}{|c|c|c|c|c|} \hline
 $\quad$Method$\quad$& $Q_{min}^2,\,\GeV^2$ & $\quad\mu_4/M^2\quad$ &
 $\quad\mu_6/M^4\quad$ & $\quad\mu_8/M^6\quad$  \\ \hline\hline
                     &  0.50 & $-0.043(3)$  &     0        &    0       \\
   NLO PDG           &  0.30 & $-0.074(3)$  & $~ 0.026(7)$ &    0       \\
                     &  0.27 & $-0.049(4)$  & $-0.010(3)$  &  0.010(1)  \\ \hline\hline
                     &  0.47 & $-0.049(3)$  &     0        &    0       \\
   NLO Denom     &  0.17 & $-0.069(4)$  &  0.014(1)    &    0       \\
                     &  0.17 & $-0.065(7)$  &  0.011(3)    &  0.0003(7)  \\ \hline\hline
                     &  0.47 & $-0.061(3)$  &     0        &    0       \\
   NLO SGF           &  0.19 & $-0.073(3)$  & 0.010(3)     &    0       \\
                     &  0.10 & $-0.077(4)$  & 0.014(5)     & $-0.0008(3)$  \\ \hline\hline
                     &  0.47 & $-0.055(3)$  &     0        &    0        \\
   NNLO APT          &  0.17 & $-0.062(4)$  &  0.008(2)    &    0        \\
   no evolution      &  0.10 & $-0.068(4)$  &  0.010(3)    & $-0.0007(3)$  \\ \hline
                     &  0.47 & $-0.051(3)$  &     0        &    0        \\
   NNLO APT          &  0.17 & $-0.056(4)$  &  0.0087(4)   &    0        \\
   with evolution    &  0.10 & $-0.058(4)$  &  0.0114(6)   & $-0.0005(8)$  \\ \hline
\end{tabular} \end{center}
\end{table}

In Fig.~\ref{fig:Bjtot}, we show best fits of the combined data set
for the BSR function $\Gamma_1^{p-n}(Q^2)$ with NLO Denom (solid
lines) and PDG (dashed lines) couplings and NNLO APT (dash-dotted
lines) at fixed $\Lambda_{QCD}$ value corresponding to the world
average. We also show here the pQCD part of the BSR at different
values of $\Lambda_{QCD}=300,\,400,\,500$ MeV calculated within APT
(short-dashed lines) and the SGF model \cite{Simonov} at different
values of the glueball mass $M_0=1.2,\,1.0,\,0.8\,\GeV$ (with
$\Lambda=360$ MeV) (dotted lines).

The corresponding numerical results are given in
Table~\ref{tab:Bjtot}. As we have seen before in
Fig.~\ref{fig:as-2L}, the behavior of SGF and APT couplings is very
similar in the low-energy domain $\Lambda_{QCD}<Q\lesssim 1$ GeV. As
a result, the corresponding perturbative parts of BSR in
Fig.~\ref{fig:Bjtot} and results for higher-twist terms in
Table~\ref{tab:Bjtot} turn out to be close, too. Our fits in APT and
the SGF model give the HT values indicating a better convergence of
the OPE series due to decreasing magnitudes and alternating signs of
consecutive terms, in contrast to the usual PT fit results.

As is seen from Table~\ref{tab:Bjtot}, there is some sensitivity of
fitted values of $\mu_4$ with respect to $Q_{min}$ variations;
namely, it increases in magnitude when one incorporates into the fit
the data points at lower energies. This property of the fit may be
treated as the slow (logarithmic) evolution $\mu_4(Q^2)$ with $Q^2$
which becomes more noticeable at broader fitting ranges in $Q^2$, as
discussed above. So for completeness we included in
Table~\ref{tab:Bjtot} APT fits for $\mu_4(Q_0^2)$ taking into
account their RG evolution with $Q_0=1\,\GeV$ as a normalization
point. We see that the fit results become more stable with respect
to $Q_{min}$ variations.

However, there is still a problem with how to treat the evolution of
higher-twist terms $\mu_{6,8,..}(Q^2)$ which again may turn out to
be important when one goes to lower $Q^2$, since the fit becomes
more sensitive to very small variations of $\mu_{6,8,..}$ with
$Q^2$.

Note that the APT functions $\ac_k$ contain the $(Q^2)^{-k}\,$ power
contributions which effectively change the fitted values of $\mu$
terms. In particular, subtracting an extra $(Q^2)^{-1}$ term induced
by the APT series
\begin{eqnarray*}
&&\Gamma^{p-n}_{1,APT}(Q^2)\simeq\frac{g_A}{6}+
f\biggl(\frac{1}{\ln(Q^2/{\Lambda^{(1)}_{eff}}^2)}\biggr)+
\varkappa\frac{{\Lambda^{(1)}_{eff}}^2}{Q^2}+ {\cal
O}\left(
\frac{1}{Q^4}
\right)
\end{eqnarray*}
with $\varkappa=0.43$ and using the value
$\mu_{4,APT}^{p-n}/M^2=-0.058$ (with evolution) from
Table~\ref{tab:Bjtot}, we finally get
\begin{eqnarray}                                
 \label{mu4_APT}
 \frac{\mu_{4,APT}^{p-n}+\varkappa{\Lambda^{(1)}_{eff}}^2}{M^2}\simeq
 \frac{\mu^{p-n}_4(1\,\GeV^2)}{M^2}\simeq-0.042\,,\quad
 \Lambda^{(1)}_{eff}\sim0.18\,\GeV\,
\end{eqnarray}
that nicely correlates with the result in Ref.~\cite{Bj08-tw}:
$\mu^{p-n}_4/M^2\simeq-0.045.$ This demonstrates the concert of the
APT analysis with the usual PT one for the BSR data at
$Q^2\geq1\;\GeV^2$.

We do not take into account RG evolution in $\mu_4$ for the standard
PT calculations since the only effect of that would be the
enhancement of the Landau singularities by extra divergencies at
$\Lambda_{QCD}$ (see Fig.~\ref{fig:evol-mu4}), whereas at higher
$Q^2\sim 1\,\GeV^2$ the evolution is negligible with respect to
other uncertainties. In ghost-free models, however, the evolution
gives a noticeable effect at low $Q\sim\Lambda_{QCD}$. Note that our
previous result in Ref.~\cite{Bjour}, obtained without taking into
account the RG evolution, turned out to be slightly larger than
(\ref{mu4_APT}) $\mu^{p-n}_4/M^2\simeq-0.048$ which is very close to
the corresponding value obtained with the most precise Denom PT
coupling and is shown in Table~\ref{tab:Bjtot}.

\subsection{Singlet case: spin sum rules $\Gamma_1^{p,n}$ and nucleon spin structure}

Turn now to the three-loop APT part of the proton moment
$\Gamma^p_{1,APT}(Q^2)$. Its value is quite stable with respect to
small variations of $\Lambda$, in contrast to the huge instability
of $\Gamma^p_{1,PT}$: it changes now by about $2\%-3\%$ within the
interval $\Lambda^{(3)}=300-500\,\MeV\,$. The same was previously
observed for the Bjorken function $\Gamma^{p-n}_{1,APT}(Q^2)$ in
Ref.~\cite{Bjour}. Because of this fact the low-$Q^2$ data on
$\Gamma^p_1(Q^2)$ cannot be used for determination of $\Lambda$ in
the APT approach.

Extending the analysis of Ref.~\cite{MSS} to lower $Q^2$ scales, we
estimated the relative size of APT contributions to
$\Gamma_1^p(Q^2)$. It turned out that the third term $\sim\ac_3$
contributes no more than $5\%$ to the sum, thus supporting the
practical convergence of the APT series.

\begin{table}[!h]
\caption{\small\sf Sensitivity of the best APT fit results of proton
$\Gamma^p_1(Q^2)$ data (elastic contribution excluded) to
$\Lambda_{n_f=3}$ variations. The minimal fitting border is
$Q_{min}^2=0.12\,\GeV^2$.}
\begin{center}\label{tab:Lam-APT}
\begin{tabular}{|c|c|c|c|c|} \hline
 $\;\Lambda_{QCD},\,\MeV\;$& $\qquad a^{inv}_0\qquad$ &
 $\quad\mu_4/M^2\quad$ & $\quad\mu_6/M^4\quad$ & $\quad\mu_8/M^6\quad$
 \tabularnewline \hline\hline
 300 & 0.43(3) & $-0.082(4)$  &  0.015(9) & $-0.0009(5)$ \tabularnewline
 400 & 0.45(3) & $-0.081(4)$  &  0.015(9) & $-0.0009(5)$ \tabularnewline
 500 & 0.47(3) & $-0.080(4)$  &  0.014(9) & $-0.0009(5)$
 \tabularnewline \hline
\end{tabular} \end{center}
\end{table}

To see how the numerical fit results are sensitive to
$\Lambda_{(n_f=3)}$ in APT, we fulfilled four different fits of the
proton $\Gamma^p_1(Q^2)$ data with
$\Lambda_{QCD}=300,\,400,\,500\,\MeV$ as we did before in the
standard PT. The results of these fits are shown in
Table~\ref{tab:Lam-APT}. Comparing these results with the data from
Table~\ref{tab:Lam}, we see that the corresponding results in the
standard PT are much more sensitive to $\Lambda$ variations than
ones in APT.

\begin{figure}[!h]
 \centerline{\epsfig{file=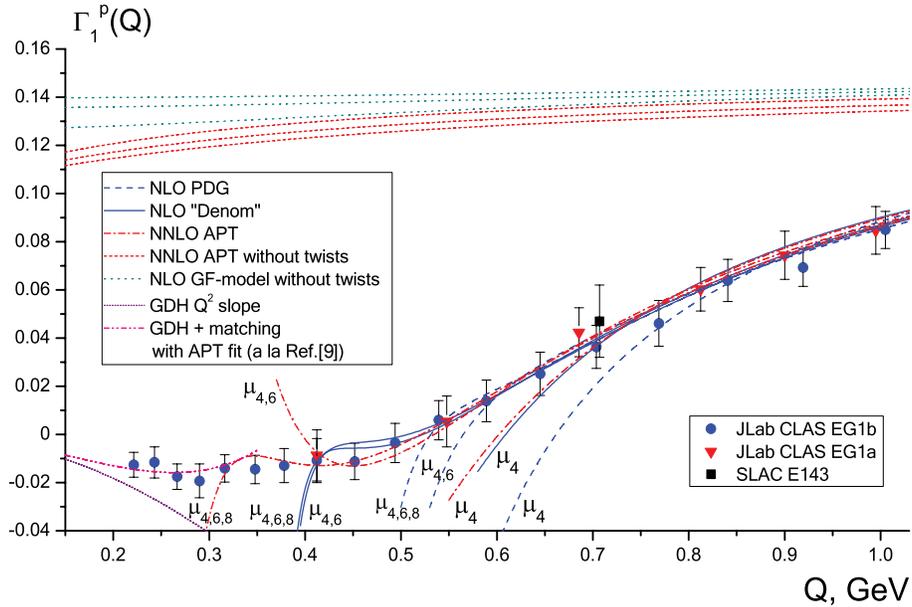,height=8cm,width=12cm}}
 \caption{\footnotesize Best (1,2,3+1)-parametric fits of the JLab and SLAC
 data on $\Gamma^p_1$ (elastic contribution excluded).}
 \label{fig:prottot}
\end{figure}

In Fig.~\ref{fig:prottot}, we show best fits of the combined data
set for the function $\Gamma_1^{p}(Q^2)$ (the data uncertainties are
statistical only) in the standard PT (PDG and Denom versions) and
the APT approaches. We have also shown the perturbative parts of
$\Gamma_1^p(Q^2)$ calculated in APT and the SGF model. They are
close to each other down to $Q\sim\Lambda$, similar to the BSR
analysis in the previous subsection. A similar observation was made
in the analysis of the small $x$ spin averaged structure functions
in Ref.~\cite{Kotikov}.

In Table~\ref{tab:prottot}, we present the combined fit results of
the proton $\Gamma^p_1(Q^2)$ data (elastic contribution excluded) in
APT, the SGF model and conventional PT in PDG and denominator forms.
One can see there is noticeable sensitivity of the extracted
$a^{inv}_0$ and $\mu_4$ with respect to the minimal fitting scale
$Q^2_{min}$ variations, which may be (at least, partially)
compensated by their RG $\log Q^2$ evolution, similar to the BSR
case. For completeness we included in Table~\ref{tab:prottot} APT
fits for $a^{inv}_0(Q_0^2)$ and $\mu_4(Q_0^2)$, taking into account
their RG evolution.

\begin{table}[!h]
\caption{\small\sf Combined fit results of the proton
$\Gamma^p_1(Q^2)$ data (elastic contribution excluded). APT fit
results $a_0$ and $\mu^{APT}_{4,6,8}$ (at the scale
$Q_0^2=1\,\GeV^2$) are given without and with taking into account
the RG $Q^2$ evolution of $a_0(Q^2)$ and $\mu^{APT}_4(Q^2)$.}
\begin{center}\label{tab:prottot}
\begin{tabular}{|c|c|c|c|c|c|} \hline
 $\quad$Method$\quad$& $Q_{min}^2,\,\GeV^2$ & $\qquad a_0\qquad$ &
 $\quad\mu_4/M^2\quad$ & $\quad\mu_6/M^4\quad$ & $\quad\mu_8/M^6\quad$  \\ \hline\hline
                     &  0.59 & 0.33(3) & $-0.050(4)$  &     0     &    0       \\
   NLO PDG           &  0.35 & 0.43(5) & $-0.087(9)$  &  0.024(5) &    0       \\
                     &  0.29 & 0.37(5) & $-0.060(15)$ &  -0.001(8)&  0.006(5)  \\ \hline\hline
                     &  0.59 & 0.35(3) & $-0.058(4)$  &     0     &    0       \\
   NLO Denom     &  0.20 & 0.38(3) & $-0.076(4)$  &  0.013(1) &    0       \\
                     &  0.17 & 0.38(4) & $-0.070(8)$  &  0.010(4) &  0.0004(5)  \\ \hline\hline
                     &  0.47 & 0.32(4) & $-0.056(4)$  &     0     &     0       \\
   NLO SGF           &  0.17 & 0.36(3) & $-0.071(4)$  & 0.0082(9) &    0       \\
   $M_0=1\,\GeV$     &  0.10 & 0.40(4) & $-0.080(4)$  & 0.0134(9) & $-0.0007(6)$  \\ \hline\hline
                     &  0.47 & 0.35(4) & $-0.054(4)$  &     0      &    0        \\
   NNLO APT          &  0.17 & 0.39(3) & $-0.069(4)$  &  0.0081(8) &    0        \\
   no evolution      &  0.10 & 0.43(3) & $-0.078(4)$  &  0.0132(9) & $-0.0007(5)$  \\ \hline
                     &  0.47 & 0.33(4) & $-0.051(4)$  &     0      &    0        \\
   NNLO APT          &  0.17 & 0.31(3) & $-0.059(4)$  &  0.0098(8) &    0        \\
   with evolution    &  0.10 & 0.32(4) & $-0.065(4)$  &  0.0146(9) & $-0.0006(5)$  \\ \hline
\end{tabular} \end{center}
\end{table}

As we already mentioned, the evolution of the $\mu^p_4(Q^2)$ is
taken to be the same as for the nonsinglet term $\mu^{p-n}_4(Q^2)$,
allowing one to keep only one fitting parameter $\mu^p_4(Q_0^2)$
instead of two in the general case. We also tested that the singlet
anomalous dimension instead of the nonsinglet one [resulting in the
same $Q^2$ evolution of $\mu^p_4(Q^2)$ as that of
$\mu^{p+n}_4(Q^2)$] leads to close fit results within error bars.

\begin{figure}[!h]
\begin{center}
\epsfig{file=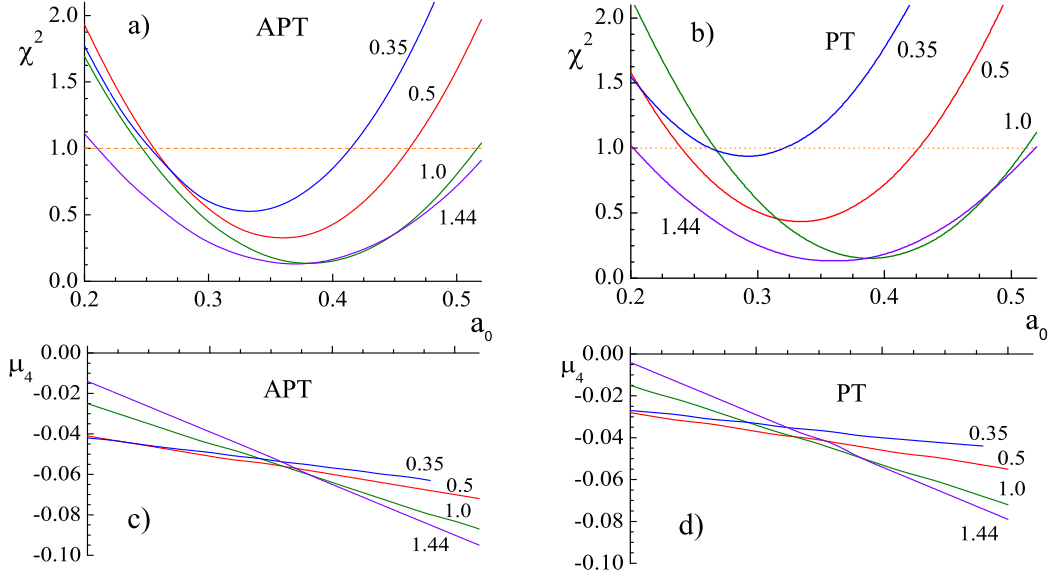,width=5.5in} \vspace{0.1in} \caption
{Behavior of $\chi^2/D.o.f.$ and $\mu^p_4$ from the proton data fits
(with only one $1/Q^2$ term) as functions of $a_0$ at different
values of $Q_{min}^2$ (the numbers at the curves) in the APT (left
panels) and PT (right panels) cases.} \label{fig:a0-chi2-mu4}
\end{center}
\end{figure}

Figure~\ref{fig:a0-chi2-mu4} demonstrates the characteristic values
of the proton data fits $\chi^2/D.o.f.$ (upper row) and the twist-4
coefficient $\mu_4$ (lower row) as functions of $a_0$ at different
values of $Q_{min}^2$ (numbers at the curves). One can see that at
lower $Q^2$ ($Q_{min}^2 < 1 ~\GeV^2$) the APT description (left
panels) turns out to be more precise and stable than that in the
standard PT (right panels). Though we have taken the fitted values
of $a_0$ and higher twists $\mu_{2i}$ in the minima of each
$\chi^2/D.o.f.$ curve as best fits, the naive constraint
$\chi^2/D.o.f.\leq1$ (dotted horizontal lines mark $1$) provides a
quite wide spread in the allowable values of the fit parameters.
However, it would be reasonable to take the spread between different
minima as an optimistic error bar of our analysis. This gives us the
following result: $a_0=0.33\pm0.05$, which is consistent with the
recent analysis by COMPASS \cite{COMPASS06} and HERMES
\cite{HERMES06} (see Table~\ref{table1}).

\begin{figure}[!h]
 \centerline{\epsfig{file=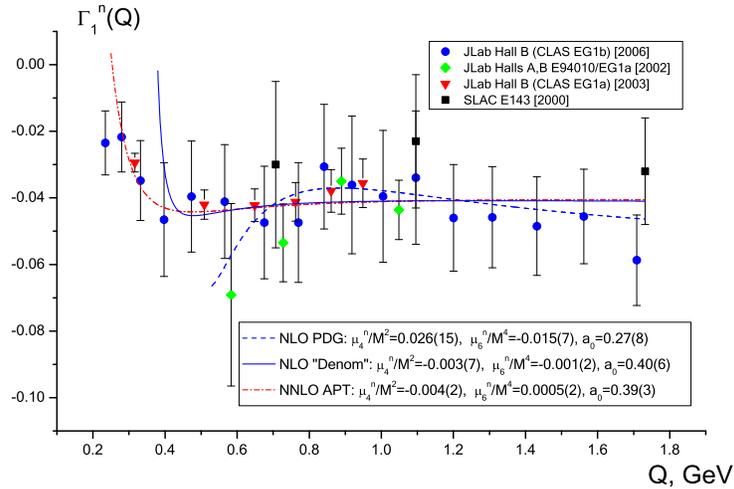,height=7cm,width=10cm}}
 \caption{\footnotesize Best (2+1)-parametric fits of the JLab and SLAC
 data on $\Gamma^n_1$ calculated with NLO Denom (solid line) and PDG
 (dashed line) couplings and NNLO APT (dash-dotted line).}
 \label{fig:neutron}
\end{figure}

In Fig.~\ref{fig:neutron}, we show the best fit results for the less
precise neutron $\Gamma^n_1(Q^2)$ data. Again, the APT fit gives the
HT values demonstrating a better convergence of the OPE series, in
contrast to the usual PT fit results. Fits with APT and more precise
Denom PT couplings lead to a much smaller value of $\mu^n_4$ and
more stable fitting curves than that with the PDG coupling. Also the
axial singlet charge $a_0$ extracted within APT from the neutron
data turns out to be very close to the one extracted from more
precise proton data (see Table~\ref{tab:prottot}).

To obtain the genuine value of the twist-4 term $\mu_4^p$, we act in
a similar way as for the BSR case in the previous subsection, namely,
subtracting an extra $(Q^2)^{-1}$ term induced by the APT series
\begin{eqnarray}\nonumber
E^{APT}_{NS}(Q^2)&=&E_{NS}(\alps=\alps^{LO}(Q^2))+
\varkappa^{NS}_4 \, \frac{{\Lambda^{(1)}_{eff}}^2}{Q^2}+
{\cal O}\left(\frac{1}{Q^4}\right), \\[0.2cm]
E^{APT}_{S}(Q^2)&=&E_{S}(\alps=\alps^{LO}(Q^2))+
\varkappa^{S}_4 \, \frac{{\Lambda^{(1)}_{eff}}^2}{Q^2}+{\cal
O}\left(\frac{1}{Q^4}\right) \label{Q-n}
\end{eqnarray}
with $\Lambda^{(1)}_{eff}\sim0.18\,\GeV,\,\varkappa^{NS}_4=2.035$,
and $\varkappa^{S}_4=0.661$, and using the fit result in APT (with
evolution) $\mu_4^{p,APT}/M^2=-0.065$ from Table~\ref{tab:prottot},
we obtain
\begin{eqnarray}
\label{mu4_APT-p}
 \frac{\mu^p_4(1\,\GeV^2)}{M^2}\simeq\frac{1}{M^2}\left(\mu_4^{p,APT}+\frac{1}{12}
 \biggl(a_3+\frac13a_8\biggr)\varkappa^{NS}_4{\Lambda^{(1)}_{eff}}^2+\frac{1}{9}
a^{inv}_0\,\varkappa^{S}_4{\Lambda^{(1)}_{eff}}^2\right)\simeq-0.055\,.
\end{eqnarray}

Analogously, for a neutron we have $\mu^n_4/M^2\simeq-0.010.$
Subtracting it from the proton value (\ref{mu4_APT-p}), we get for
the nonsinglet twist-4 term $\mu^{p-n}_4/M^2\simeq-0.045\,,$ which
is close to the result in Ref.~\cite{Bj04-tw}, showing up the
consistence of the APT analysis with the usual PT one for the proton
and neutron SSR $\Gamma^{p,n}_1$ data at $Q^2\geq1\;\GeV^2$. Our
result (\ref{mu4_APT-p}) is also consistent with the previous
extraction at higher energies in Ref.~\cite{proton} within the error
bars (see also Table~\ref{table2}).

It is worth noting that the best APT fit allows one to describe
low-energy JLab data on $\Gamma^{p,n}_1$ at scales down to $Q\sim
350\,\MeV$ with only the first three terms of the OPE series, unlike
the usual PT case, where such fits happened to be impossible (due to
the ghost issue) even for an increasing number of HT terms. This
means that the lower bound of the pQCD applicability (supported by
power HT terms) now may be shifted down to $Q\sim\Lambda_{\rm
QCD}\simeq 350$ MeV.

However, it seems to be difficult to get a description in the region
$Q<\Lambda_{\rm QCD} $. This is not surprising, because the
expansion in positive powers of $Q^2$ and its matching \cite{SofTer}
with the HT expansion are relevant here. In this respect, the
$\Lambda_{\rm QCD} $ scale appears as a natural border between
``higher-twist'' and ``chiral'' nonperturbative physics.
\begin{figure}[!ht]
 \centerline{\epsfig{file=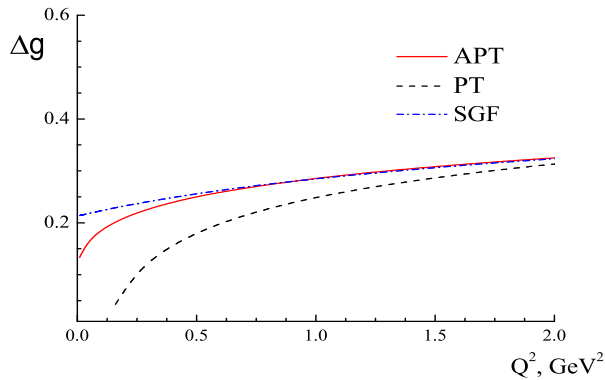,height=5cm,width=8cm}}
 \caption{\footnotesize Scale dependence of the gluon polarization $\Delta g$,
 obtained for different versions of perturbation theory -- in APT (solid line),
 in conventional PT (dashed line), and in the SGF model (dash-dotted line).}
 \label{fig:deltag}
\end{figure}

Finally, in Fig.~{\ref{fig:deltag}}, we show the scale dependence of
the gluon polarization $\Delta g$ obtained in APT, PT, and the SGF
model. In conventional PT the value of $\Delta g$ is small at the
lower scale $Q^2\sim 0.3\,\GeV^2$ (see
Ref.~\cite{Wakamatsu:2007ar}). However, as one can see from
Fig.~{\ref{fig:deltag}}, one may evolve $\Delta g$ starting from
higher scales $Q^2>1\,\GeV^2$ down to the deep infrared region and
observe that the smallness of $\Delta g$ is a consequence of the
Landau singularities in $\as$. Applying different ghost-free models
we see that $\Delta g$ is much higher at $Q^2\lesssim 0.5\,\GeV^2$
than one predicted in the standard PT.

\section{Conclusion and Outlook}

The singlet axial charge $a_0$ is the essential element of the
nucleon spin structure which is related to the average total quark
polarization in the nucleon. In this paper, we systematically
extracted this quantity from very accurate JLab data on the first
moments of spin structure functions $g^{p,n}_1$.

These data were obtained at low $Q^2$ region $0.05<Q^2<3\,{\rm
GeV}^2$, and therefore, a special attention was paid to the QCD
coupling in this domain. We demonstrated that the denominator form
(\ref{Denom}) of the QCD coupling \as is more suitable at the low
$Q^2$ (see Figs.~\ref{fig:as-2L} and \ref{fig:loop-sen}). In
particular, at the two-loop level it happens to be quite close to
the exact numerical solution of the corresponding two-loop RG
equation for $Q\gtrsim0.5\,\GeV$.

The performed analysis includes even lower $Q\sim\Lambda_{QCD}$ and
involves the QCD coupling which is free of Landau singularities. For
this purpose we used the APT \cite{apt96-7} and the soft
glueball-freezing model \cite{Simonov} for the infrared-finite QCD
coupling $\as$. It was shown that the singularity-free APT and SGF
QCD couplings are very close in the domain $Q\gtrsim 400\,\MeV$.

One can argue that large order perturbative and nonperturbative
contributions are mixed up, and the duality between them is expected
(see Ref.~\cite{ZN09}). We tested a separation of perturbative and
nonperturbative physics and performed a systematic comparison of the
extracted values of the higher-twist terms in different versions of
perturbation theory. A kind of duality between higher orders of PT
and HT terms is observed so that higher order terms absorb part of
the HT contributions moving the pQCD frontier between the PT and HT
contribution to lower $Q$ values in both nonsinglet and singlet
channels (see Fig.~\ref{fig:dual}). As expected, the value of $a_0$
changes substantially when coming from LO to NLO, whereas it is
quite stable in higher-loop approximations.

The perturbative contribution to the proton spin sum rule
$\Gamma^p_1$ and to the Bjorken sum rule $\Gamma^{p-n}_1$ in the APT
approach and the SGF model is less than 5 \% for $Q>\Lambda$. This
explains the similarity of the extracted higher-twist parameters for
these two modifications of QCD couplings.

In the APT approach the convergence of both the higher orders and HT
series is much better. In both the nonsinglet and singlet case,
while the twist-4 term happened to be larger in magnitude in the APT
than in the conventional PT, the subsequent terms are essentially
smaller and quickly decreasing (as the APT absorbs some part of
nonperturbative dynamics described by HT). This is the main reason
for the shift of the pQCD frontier to lower $Q$ values. A
satisfactory description of the proton SSR and BSR data down to
$Q\sim\Lambda_{QCD}\simeq 350\,\MeV\,$ was achieved by taking the
higher-twist and (analytic) higher order perturbative contributions
into account simultaneously (see Figs.~\ref{fig:Bjtot} and
\ref{fig:prottot}). The best accuracy for the extracted values of
$a_0$ and higher-twist contributions $\mu_{2i}$ is achieved for the
most precise proton SSR data while the analysis of the data on the
neutron SSR shows the compatibility with the analysis of the BSR
which is free from the singlet contribution.

For the first time we considered the QCD evolution at low $Q^2$ of
both the leading twist $a_0$ and the higher-twist $\mu_4$ terms
using the (fractional) analytic perturbation theory \cite{Bakulev}
and also the related evolution of the average gluon polarization
$\Delta g$. Account of this evolution, which is most important at
low $Q^2$, improves the stability of the extracted parameters whose
$Q^2$ dependence diminishes (see Table~\ref{tab:prottot}). As a
result, we extract the value of the singlet axial charge
$a_0(1\,\GeV^2)=0.33\pm0.05$. This value is very close to the
corresponding COMPASS $0.35\pm 0.06$ \cite{COMPASS06} and HERMES
$0.35\pm 0.06$ \cite{HERMES06} results.

The RG evolution of $a_0$ is related to the evolution of the average
gluon polarization $\Delta g$ \cite{Anselmino:1994gn,Leader08}. The
results of the evolution of $\Delta g$ in the analytic perturbation
theory and in the standard PT was compared (see
Fig.~\ref{fig:deltag}). The decrease of $\Delta g$ at low $Q^2$ in
APT is not so dramatic as in the standard PT case
\cite{Wakamatsu:2007ar}.

In a sense, it could be natural if the main reason for the
significant shift of the pQCD frontier to lower $Q^2$ scales was the
disappearance of unphysical singularities in perturbative series.
Note that the data at very low $Q\sim\Lambda_{QCD}$ are usually
dropped from the analysis of $a_0$ and the higher-twist term in the
standard PT analysis because of Landau singularities. At the same
time, the compatibility of our results for $a_0$, extracted from the
low energy JLab data with previous results \cite{COMPASS06,HERMES06}
demonstrates the universality of the nucleon spin structure at large
and low $Q^2$ scales. It will be very interesting to explore the
interplay between perturbative and nonperturbative physics against
other low energy experimental data.

\section*{ACKNOWLEDGMENTS}
This work was partially supported by RFBR Grants No. 07-02-91557,
No. 08-01-00686, No. 08-02-00896-a, and No 09-02-66732, the
JINR-Belorussian Grant (Contract No. F08D-001), and RF Scientific
School Grant No. 1027.2008.2. We are thankful to A.P.~Bakulev,
J.P.~Chen, G.~Dodge, A.E.~Dorokhov, S.B.~Gerasimov, G.~Ingelman,
A.L.~Kataev, S.V.~Mikhailov, A.V.~Sidorov, D.B.~Stamenov, and
N.G.~Stefanis for valuable discussions.

\end{document}